\documentclass[twocolumn, trackchanges]{aastex7}

\usepackage{amsmath}     
\usepackage{subcaption}  
\usepackage{CJK}         

\graphicspath{
{./}
{figures/opacity_laws/}
{figures/typical_SEDs/}
{figures/paras_survey/}
{figures/stacked_SEDs/}
{figures/summary/}
{figures/PRIMA/}
}

\newcommand{\msun}{M_\odot}

\newcommand{\zsun}{Z_\odot}

\newcommand{\cc}{{\rm cm}^{-3}}

\newcommand{\pc}{{\rm pc}}
\newcommand{\mum}{\mu {\rm m}}
\newcommand{\kms}{{\rm km~s}^{-1}}

\newcommand{\K}{{\rm K}}
\newcommand{\beq}{\begin{equation}}
\newcommand{\eeq}{\end{equation}}
\newcommand{\muJy}{\mu{\rm Jy}}

\newcommand{\Lnureemit}{L_\nu^\text{re-emit}}
\newcommand{\Lnuinc}{L_\nu^\text{inc}}
\newcommand{\Lnuobs}{L_\nu^\text{obs}}
\newcommand{\NH}{N_{\rm H}}
\newcommand{\AVlim}{A_V^{\rm lim}}

\newcommand{\blue}[1]{ #1}

\begin{document}
\begin{CJK*}{UTF8}{gbsn}

\title{Dust Budget Crisis in Little Red Dots}

\author[orcid=0009-0005-1831-3042,sname='']{Kejian Chen (陈可鉴)}
\affiliation{Kavli Institute for Astronomy and Astrophysics, Peking University, Beijing 100871, China}
\affiliation{Department of Astronomy, School of Physics, Peking University, Beijing 100871, China}
\email[show]{chenkejian@stu.pku.edu.cn}

\author[orcid=0000-0002-8502-7573,sname='']{Zhengrong Li}
\affiliation{Kavli Institute for Astronomy and Astrophysics, Peking University, Beijing 100871, China}
\affiliation{Department of Astronomy, School of Physics, Peking University, Beijing 100871, China}
\email[]{lizhengrong@pku.edu.cn}  

\author[orcid=0000-0001-9840-4959,sname='']{Kohei Inayoshi}
\affiliation{Kavli Institute for Astronomy and Astrophysics, Peking University, Beijing 100871, China}
\email[show]{inayoshi@pku.edu.cn}

\author[orcid=0000-0001-9840-4959,sname='']{Luis C. Ho}
\affiliation{Kavli Institute for Astronomy and Astrophysics, Peking University, Beijing 100871, China}
\affiliation{Department of Astronomy, School of Physics, Peking University, Beijing 100871, China}
\email[]{}

\begin{abstract}
Little red dots (LRDs), a population of active galactic nuclei (AGNs) recently identified by JWST, are characterized by their compact morphology and red optical continuum emission, which is often interpreted as evidence for significant dust extinction of $A_V \gtrsim 3$ mag. 
However, the dust-reddened AGN scenario is increasingly challenged by their faint near-to-far infrared emission and a potential ``dust budget crisis" in cases when the host galaxy is either undetectably low-mass or absent. 
In this study, we re-evaluate the dust extinction level in LRDs by modeling the UV-to-infrared spectra for various extinction laws and a broad range of dusty distribution parameters.
Comparing the predicted infrared fluxes with observational data from the JWST MIRI, Herschel, and ALMA,
our analysis finds that the visual extinction is \blue{tightly and consistently constrained to $A_V \lesssim 1.0-1.5~{\rm mag}$ for A2744-45924, RUBIES-BLAGN-1, 
and stacked SEDs from a large sample of LRDs} under the SMC extinction law, with slightly weaker constraints for those with gray extinction in the UV range. 
The revised $A_V$ values yield radiative efficiencies of $\sim 10\%$ for the LRD population, easing the tension with the So{\l}tan argument for the bulk AGN population at lower redshifts.
Moreover, this moderate extinction (or dust-free) scenario, with reprocessed emission spectra testable by future far-infrared observatories,
provides a paradigm shift in understanding their natures, environments, and evolutionary pathways of massive black holes in the early universe.
\end{abstract}


\section{Introduction} 
\end{CJK*}

The James Webb Space Telescope (JWST) has uncovered a new type of active galactic nuclei (AGNs), known as little red dots (LRDs), characterized by broad emission lines, red continuum spectra, and compact morphology \citep{Kocevski_2023,Matthee_2024,Greene_2024,Kokorev_2023_aLRDwNIRSpec,Wang_2024_LRDGalaxy,Labbe_2025_LRD_ALMA}.
This new population is $\sim 1-2$ orders of magnitude more abundant than luminous quasars, reaching $\gtrsim 1\%$ of the abundance of star-forming galaxies at redshifts of $z\sim 4-7$ \citep{Kocevski_2024,Kokorev_2024_LRDcensus,Akins_COSMOS-Web_2025}, and is considered to represent a key building block and crucial growth phase of seed black holes (BHs) that power quasars at later epochs \citep{Fujimoto_2022}.
Moreover, the population emerges at high redshifts and rapidly disappears toward lower redshifts \citep{Kocevski_2024,Ma_2025_Local_LRDs,Euclid_LRD_2025}, possibly transitioning into more typical AGN populations by losing the LRD characteristics \citep{Inayoshi_2025_LRD_LogNorm}.

A well-defined feature of broad-line AGNs observed as LRDs is their characteristic v-shaped spectral energy distribution (SED), composed of red optical and blue UV continua emission with a turnover wavelength near the Balmer limit \citep{Kocevski_2023,Barro_2024,Matthee_2024,Greene_2024,Hainline_2025,Setton_2025a}.
Although the origin of this distinctive SED has been poorly understood, the red component is often interpreted as dust-reddened systems with extinction levels of $A_V\sim 3~{\rm mag}$,
suggesting scenarios involving dusty star-forming galaxies, AGNs, or some combination of both \citep{Wang_2024_LRDGalaxy}.
However, JWST MIRI observations have revealed that the rest-frame near-infrared (NIR) fluxes of LRDs tend to be fainter than expected from hot dust heated by UV/optical radiation sources \citep{Perez-Gonzalez_2024_MIRILRD,Akins_COSMOS-Web_2025,Leung_2024_LRDPRIMERMIRI,Setton_2025b}.
To satisfy these MIRI flux constraints (or upper limit if non-detection is reported), dust distribution should be either spatially extended \citep{LiZ_2025_LRDDustSED}, unlike the compact dusty tori commonly found in AGNs, or highly clumpy (\citealt{Nenkova_2008,Honig_2010}, and its application to the LRD context in \citealt{Casey_2024_DustInLRD}).
Both the scenarios effectively shift the dust emission peak from the NIR ($\simeq 1-5~\mum$) to the mid-infrared (MIR; $\simeq 5-35~\mum$) bands.
Furthermore, non-detection of LRDs in ALMA suggests that dust is as warm as $\sim 100~\K$, substantially higher than that led by active star formation \citep{Labbe_2025_LRD_ALMA,Akins_COSMOS-Web_2025,Xiao_2025_NoCIILRD}.
\cite{Casey_2025_DustMassUpper} also reported non-detection of stacking deep ALMA observations for 60 LRDs, providing a stringent upper limit on the dust mass below $10^6~\msun$. 

Additionally, there is no clear observational evidence for host galaxies associated with LRDs.
This might not be surprising as LRDs are typically selected as point-like sources in imaging from longer-wavelength channels of JWST NIRCam. 
In some cases, extended UV components have been detected \citep{Killi_2024_LRD,Chen_2024_HostG_LRD,Chen_nebula_2025,Rinaldi_2024}, but these features are often spatially offset 
from the central LRD or show surface brightness profiles with low central concentration. 
For the underlying host galaxies, if present, stringent upper limits on the stellar mass are placed at $\log (M_\star/\msun) < 8.3-9.6$ based on imaging decomposition and spectral fitting \citep{Chen_2024_HostG_LRD}\footnote{\blue{\citet{Kokorev_2024_absorptionlines} have reported the detection of Balmer absorption features on the spectrum of an LRD, which are interpreted as absorption lines originated from evolved stellar populations. However, such spectral features are rarely observed and might be non-stellar but AGN origin \citep{Inayoshi_Maiolino_2025}.}}.
If the LRD population indeed lacks stellar hosts, or contains only minimal stellar component, this raises a critical question: 
{\it where do the metals and dust grains responsible for reddening the AGN originate?}
The possible absence of host galaxies highlights a potential ``dust budget crisis" for LRDs.


In this {\it Letter}, we revisit the dust-reddening hypothesis for LRD spectra and assess its necessity.
Assuming that the observed UV-to-optical spectra of LRDs result from dust-reddening with a given extinction level $A_V$, 
we reconstruct the expected IR emission from dust heated by the central radiation source.
By comparing the predicted reprocessed IR SED with current observations, we find that the upper limits on $A_V$ for two broad-line LRDs with rest-frame UV-to-IR coverage \citep{Wang_2024_z3LRDwMIRI,Labbe_2024_MostBrightLRD,Setton_2025b} and two stacked LRD samples \citep{Delvecchio_agn-heated_2025,Akins_COSMOS-Web_2025} are constrained as low as $A_V\lesssim 0.5 \text{--} 1.5~{\rm mag}$, depending on the assumed dust reddening law. 
This moderate level of dust extinction challenges the conventional interpretations of LRDs as heavily obscured AGNs, and instead motivates a paradigm shift in understanding their properties, environments, and potential connection to their descendant populations.

\section{Method}

\begin{figure}[t]
    \centering
    \includegraphics[width=1.0\linewidth]{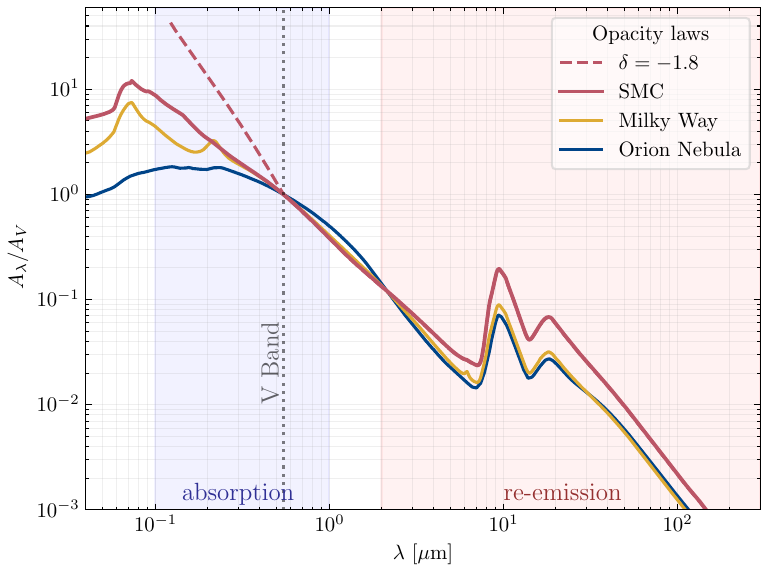}
    \caption{Extinction curves as a function of wavelength used in this work, covering from UV to infrared wavelengths: the SMC, the interstellar dust in the Milky Way with $R_V = 3.1$ \citep{WeingartnerDrain_2001_extinction}, and the Orion Nebula \citep{Baldwin1991}. We also show the modified Calzetti's attenuation law proposed in \cite{Noll_2009_flexible_dust_law} with $\delta = -1.8$ (see main text). The extinction $A_{\lambda}$ is normalized at the V band 5500~{\AA} (dotted vertical line). 
    The UV to optical part is used to compute dust attenuation of 
the incident radiation flux (blue shaded region), and the IR part is used to calculate the re-emission from heated dust (red shaded region). 
    } 
    \label{fig:opacities}
\end{figure}

\subsection{Dust Opacity}
\label{subsec:dust_opacity}

In this work, we adopt three different dust extinction laws: the Small Magellanic Cloud (SMC), the Milky Way (MW) interstellar medium (MW) dust \citep{WeingartnerDrain_2001_extinction} with $R_V = 3.1$, and the Orion Nebula dust \citep{Baldwin1991}.
Figure~\ref{fig:opacities} shows the wavelength dependence of the dust opacity for each model.
In our analysis, the UV to optical parts of the extinction curves are used to compute the attenuation of 
the incident radiation flux (blue shaded region), while the IR parts are used to estimate the re-emission from heated dust (red shaded region). 
Among the three models, the SMC extinction curve shows a steep wavelength dependence at $0.1 \lesssim \lambda/\mum \lesssim 0.5$, whereas the Orion Nebula extinction curve remains relatively flat at $\lambda \lesssim 0.3~\mum$.
The MW extinction law shows the bump feature of $2175~{\rm \AA}$ possibly from small carbonaceous grains and/or
polycyclic aromatic hydro- carbon (PAH) molecules, which is absent in the other cases likely due to destruction by 
irradiation from central hot sources and size growth in dense environments \citep[e.g.][]{Reach_2000_PAHinSMC}.
At longer wavelengths, all models converge to a single power-law behavior following the electric dipole approximation $A_\lambda \propto \lambda^{-2}$ \citep{Draine_2011}, with two prominent bumps of silicate features at $9.7~\mum$ and $18~\mum$ \citep[e.g.,][]{Hao_2007}.

In addition to the extinction curves described above, we consider a flexible dust attenuation law originally proposed in \cite{Noll_2009_flexible_dust_law} as
\begin{equation}
A_\lambda = A_V\cdot \frac{k'(\lambda)}{R'_V}\left(\frac{\lambda}{0.55~\mum}\right)^{\delta},
\label{eq:noll}
\end{equation}
where $R'_V = 4.05$ and $k'(\lambda)$ is the polynomial function provided by Equation~(4) in \cite{Calzetti_2000}.
The power-law index $\delta$ controls the slope of the attenuation curve in the UV band: 
the curve becomes steeper than the original Calzetti law when $\delta<0$.
Although \cite{Noll_2009_flexible_dust_law} initially introduced the index $\delta$ to provide moderate flexibility 
($-0.3 < \delta < 0.3$) for fitting normal star-forming galaxies, this functional form has also been used to 
explain the spectral shape of some LRDs, which shows stronger UV extinction than the SMC law (corresponding to 
$\delta\simeq -0.5$) \citep[e.g.][]{Wang_2024_z3LRDwMIRI,Ma_2025,Labbe_2024_MostBrightLRD,deGraaff_2025}.
For instance, the SEDs of A2744-45924 and RUBIES-BLAGN-1 are fitted with $\delta = -1.79$ \citep{Labbe_2024_MostBrightLRD} 
and $\delta = -1.21$ \citep{Wang_2024_z3LRDwMIRI}, respectively.
Motivated by these results, we test a case with a steep attenuation law of $\delta = -1.8$ at $0.12\leq \lambda/\mum \leq 0.55$
and revert to the SMC law at longer wavelengths (dashed curve in Figure~\ref{fig:opacities})\footnote{The shorter-wavelength part of the attenuation model with $\delta=-1.8$ is not considered because the polynomial fitting function $k'(\lambda)$ of the Calzetti law is given only at $\lambda\geq 0.12~\mum$. As discussed in Section~\ref{subsec: A_V limit}, omitting the incident radiation flux at $\lambda <  0.12~\mum$ leads to a conservative upper limit on $A_V$.}.

To compute the absolute level of extinction and re-emission of radiation, we normalize the dust opacity using the relative extinction curve, $A_\lambda/A_V$,
by defining the extinction and absorption cross sections at photon frequency $\nu (=c/\lambda)$ per hydrogen atom; 
$\sigma_\nu^{\rm ext}$ and $\sigma_\nu^{\rm abs}$, respectively. 
The extinction cross section is expressed as $\sigma_\nu^{\rm ext} = \sigma_V^{\rm ext} (A_\lambda / A_V)$.
In this work, we adopt a reference value of $\sigma_V^{\rm ext} = 3.72 \times 10^{-23}~{\rm cm}^{2}~(Z/0.1~\zsun)(\mathcal{D}_\odot/0.01)$,
where $\mathcal{D}_\odot = 0.01$ is the dust-to-gas mass ratio at solar metallicity. 
We assume the dust-to-gas ratio scales linearly with metallicity as $\mathcal{D}=\mathcal{D}_\odot (Z/\zsun)$.

Although $\sigma_\nu^{\rm abs}$ generally differs from $\sigma_\nu^{\rm ext}$ due to the contribution of scattering in the latter,
the scattering properties are not well constrained from observations.
Therefore, for simplicity, we assume $\sigma_\nu^{\rm abs} = \sigma_\nu^{\rm ext}$ throughout this study.

\subsection{Dust Re-emission}
\label{subsec:dust_reemit}

In this section, we describe our method for computing the reprocessed IR spectrum $\Lnureemit$, given 
an incident AGN spectrum $\Lnuinc$, a dust density distribution following the gas number density profile $n(r)$, 
and dust cross section per hydrogen atom $\sigma_\nu$ \citep{Barvainis_1987}. 
Our approach builds upon the method presented in \cite{LiZ_2025_LRDDustSED} by implementing various dust opacity laws and exploring a wider range of physical parameters.

We assume a power-law radial density profile, $n(r) = n_0 (r/r_{\rm in})^{-\gamma}$, defined between the inner and outer radii, $r_{\rm in}$ and $r_{\rm out}$.
\blue{The density at the inner edge $n_0$ and the power-law index $\gamma$ (ranging $0 \le \gamma \le 2$) set the normalization and concentration of the dust density profile.} 
The inner edge ($r_{\rm in}$) is determined by the dust sublimation condition, where the dust temperature equals a threshold of $T_{\rm sub} = 1500~\K$.
The sublimation temperature generally depends on the grain composition and the ambient gas
density, ranging $T_{\rm sub}\simeq 1000-2000~\K$ for $10<n_0/\cc<10^{10}$ \citep{Baskin_2018}: graphite grains have a threshold temperature higher by $\sim 300~\K$ than that of silicate grains. Within the uncertainty, the upper limit of $A_V$ changes only by $\simeq 0.1~{\rm mag}$.
Given these physical parameters, the outer radius ($r_{\rm out}$) is set by the required column density $\NH(r_{\rm out})$, 
where $\NH(r) = \int_{r_{\rm in}}^{r} n(r') {\rm d}r'$, by solving the closure relation of $A_V = 1.086~\sigma_V^{\rm ext}  \NH(r_{\rm out})$.

With this dust distribution, the dust temperature profile $T_{\rm dust} (r)$ is obtained by balancing radiative heating from the incident AGN light and cooling through dust thermal emission as
\begin{equation}
\int \frac{\Lnuinc}{4 \pi r^2}e^{-\tau_\nu (r)} \sigma_\nu^{\rm abs} \mathrm{d}\nu = 4 \pi \int B_\nu[T_\mathrm{dust}(r)] \; \sigma_\nu^{\rm abs} \mathrm{d}\nu,    
\end{equation}
where the optical depth is $\tau_\nu(r) = \sigma_\nu^{\rm ext} \NH (r) $, 
$B_\nu(T)$ is the Planck function with a dust temperature $T=T_{\rm dust}$, 
and the spectrum is integrated over the entire spectral range.
Using the temperature profile, $T_{\rm dust} (r)$, the re-emitted IR spectrum is calculated as 
\begin{equation}
\Lnureemit = \Omega \int_{r_\mathrm{in}}^{r_\mathrm{out}}  \sigma_\nu^\mathrm{abs} B_\nu[T_\mathrm{dust}(r)] \; n(r) 4 \pi r^2 {\rm d}r,    
\end{equation}
where the solid angle $\Omega$ covered by the dusty medium is taken to be $4\pi$.

Based on the density profile and $A_V$ value, the total dust mass is estimated as
\begin{equation}
M_{\rm dust} = \mathcal{D} m_p \int^{r_{\rm out}}_{r_{\rm in}}n(r)4\pi r^2 {\rm d}r,    
\label{eq:Mdust}
\end{equation}
with a dust-to-gas mass ratio $\mathcal{D}$ for metallicity $Z=0.1 Z_\odot$.

\subsection{Energy-balance Constraint on $A_V$}
\label{subsec: A_V limit}

In the previous section, we described the forward modeling approach, which begins with the incident SED and then computes the IR re-emission flux for a given set of dust properties.
\blue{However, the incident SED is not directly observable, and the nature of LRDs is still debated, from pure galaxies \citep[e.g.][]{Wang_2024_LRDGalaxy} to pure AGN \citep[e.g.][]{Kocevski_2023,LiZ_2025_LRDDustSED}, even some models without dust} \citep[e.g.][]{Naidu_2025,deGraaff_2025}. 
In this study, we thus assume that the observed SED of LRDs results from dust reddening and infer the incident SED by assigning a specific value of visual extinction $A_V$ as
\begin{equation}
\label{eqn:de-extinguish}
\Lnuinc = \Lnuobs \; 10^{0.4 (A_\lambda/A_V)A_V},
\end{equation}
where $A_\lambda/A_V$ is given by the dust opacity laws (see Figure~\ref{fig:opacities}).
\blue{
We note the intricate dependence of dust distribution range ($r_{\rm in}$ and $r_{\rm out}$) on the assumed $A_V$, since a larger $A_V$ not only requires a greater column density ($N_H$) but also implies a more luminous intrinsic flux, which in turn affects the sublimation radius ($r_{\rm in}$).
}

\begin{figure*}[t]
\centering
\includegraphics[width=\textwidth]{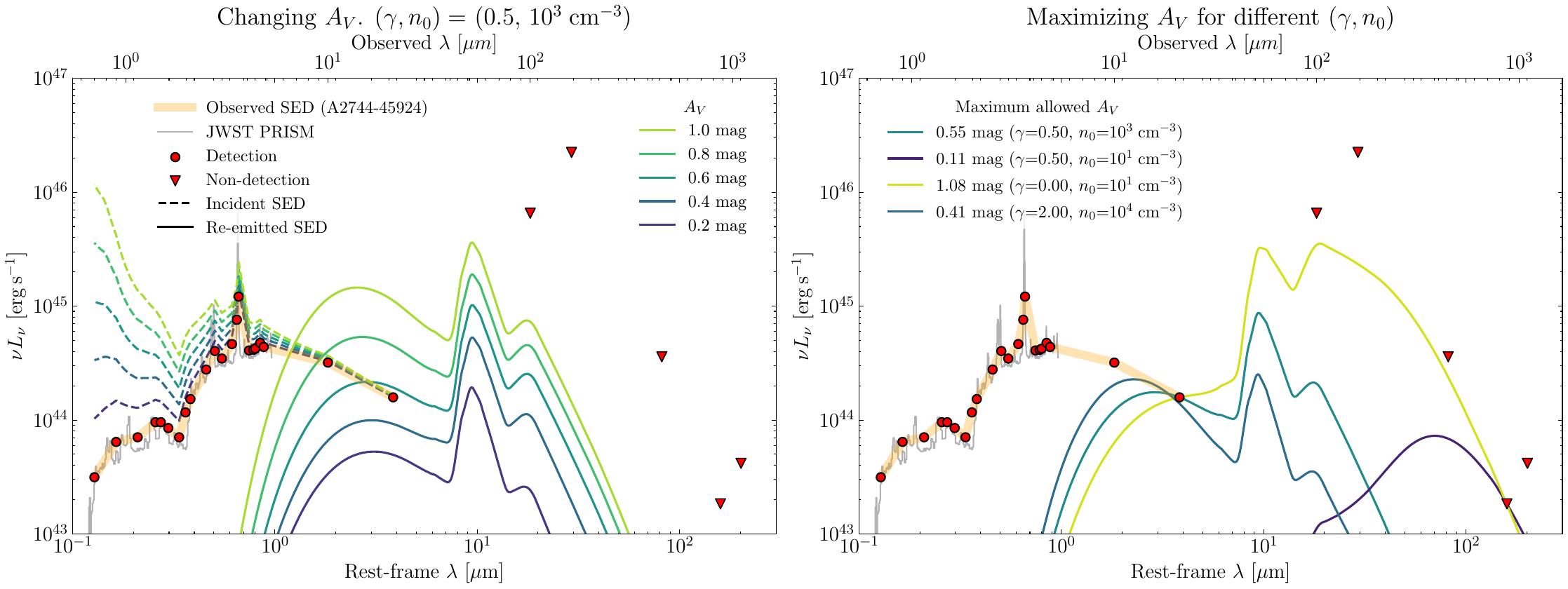}
\caption{Multi-wavelength SED of the brightest LRD, A2744-45924 (orange curve) with JWST NIRCam/MIRI detection (circles) and non-detection in Herschel and ALMA bands (triangles). {\it Left}: Incident (corrected for extinction) and reprocessed IR re-emission spectra for fixed dust distribution parameters $\gamma = 0.5$ and $n_0 = 10^3~ \cc$ with varying visual extinction $A_V$. {\it Right}: Re-emission SEDs for four different dust density configurations: $(\gamma, n_0)=(0.5,10^3~\cc)$, $(0.5,10~\cc)$, $(0,10~\cc)$, and $(2.0, 10^4~\cc)$.
For each case, the maximum allowed $A_V$ is applied such that the resulting IR flux does not violate the observational constraints.
}
\label{fig:SED with Different Av}
\vspace{5mm}
\end{figure*}

\blue{Given a photometric SED with broad wavelength coverage,} the primary criterion for deriving an upper limit on $A_V$ is that the predicted IR re-emission flux, $\Lnureemit(A_V)$, 
calculated as described in Section~\ref{subsec:dust_reemit}, must not exceed the observed flux or the upper limits from non-detections in any photometric band.
To quantify this, we define the function
\begin{equation}
f(A_V) = \max_{\nu \in \{\nu_i\} } \left[ \log_{10} \Lnureemit(A_V) - \log_{10} \Lnuobs \right],
\end{equation}
where $\{\nu_i\}$ denotes the set of photon frequencies corresponding to the pivot wavelength of photometric filters.
The \blue{maximum allowed} $A_V$ is defined by solving $f(A_V) = 0$.
We note that this constraint yields a conservative estimate of the maximum allowed $A_V$ value, since the observed IR flux should also 
include a contribution from the the dust-extinguished part of the intrinsic emission, in addition to the reprocessed component.
This method also allows us to identify which observational data provide a more stringent constraint on the $A_V$ upper limit.

It is worth noting that we do not include extreme ultraviolet (EUV) emission at rest-frame $\lambda\lesssim 0.1-0.12~\mum$
in the incident SED due to the lack of spectral data.
Although AGNs are known to be powerful ionizing radiation sources as observed in the spectra of low-redshift AGNs unaffected by intergalactic medium absorption, the intrinsic EUV properties of LRDs remain highly uncertain.
If the EUV contribution were included in our analysis, it would further heat the surrounding dusty medium and enhance 
the resulting reprocessed IR flux.
Therefore, omitting the EUV component leads to a conservative upper limit on $A_V$.

\blue{This method for constraining the dust attenuation limits is robust and generally applicable, requiring only photometric SED data (or upper limits) across the UV-to-IR wavelength range (if detailed spectral data are available, the constraint is improved). 
Therefore, this method is applicable to a large population of LRDs, and also the upcoming SED data from JWST.}

\subsection{\blue{Target SEDs for Analysis}}
\label{subsec:target_SEDs}

\blue{In this work, we apply our energy-balance analysis to four representative LRD SEDs to test the viability of the heavy dust-obscuration scenario.
The sample comprises two well-studied, individual LRDs with spectroscopic redshifts and broad emission lines detected by JWST NIRSpec, along with photometric coverage from rest-frame UV/optical (NIRCam) and NIR (MIRI).
To generalize our findings, we also analyze two stacked SEDs, which represent the average properties of the broader, photometrically-selected LRD population (spectroscopically confirmed in part; see below).}

\blue{The two individual sources are A2744-45924, the brightest LRD discovered to date, at $z_{\rm spec}=4.4655$ \citep{Labbe_2024_MostBrightLRD}, and RUBIES-BLAGN-1, a luminous LRD at $z_{\rm spec}= 3.1034$ \citep{Wang_2024_z3LRDwMIRI}.
Both exhibit clear AGN signatures (broad H$\alpha$ emission seen in the NIRCam grism; \citealt{Labbe_2024_MostBrightLRD}), but show no evidence of strong NIR or FIR dust emission, with non-detections in MIRI and ALMA suggesting a lack of both hot and cold dust components \citep{Setton_2025b}.
Moreover, the two objects have no obvious host galaxy signatures and are consistent with point-source morphology. 
In the case of A2744-45924, two extended emission components are visible, but they are spatially offset from the AGN and are consistent with foreground galaxies at $z_{\rm photo} \simeq 1$ \citep{Chen_nebula_2025}.
These two sources are selected because they are among the brightest known LRDs with the most complete multi-wavelength coverage from rest-frame UV to IR, providing the strongest observational constraints for our energy-balance models.}

\blue{For the analysis of the broader LRD population, we utilize two key, complementary stacked SED datasets from recent large-sample studies.
The first is from \citet{Delvecchio_agn-heated_2025}, based on a homogeneously-selected sample of 302 LRDs with a median redshift of $\langle z \rangle \simeq 6.2$.
A critical feature of this dataset is the significant detection in multiple JWST/MIRI bands, offering a firm anchor for the reprocessed rest-frame NIR emission.
The second is from \citet{Akins_COSMOS-Web_2025}, who stacked multi-wavelength data for 434 LRDs at $\langle z \rangle \simeq 6$.
This work provides non-detection upper limits across a wide spectral range, including relatively shallow ones in the rest-frame mid- to far-IR bands.
Analyzing these stacked SEDs allows us to assess whether the conclusions drawn from individual, bright objects hold for the average LRD population.
}

\section{Results}


\blue{In this section, we present the results from the energy-balance analysis applied to the four target SEDs described in Section~\ref{subsec:target_SEDs}. 
We first demonstrate the method and its detailed application to the individual LRDs (Sections~\ref{subsec:SED_LRD} and \ref{subsec:param_surveys}). 
We then extend the analysis to the stacked SEDs to test the robustness of our findings for the broader LRD population (Section~\ref{subsec:stacked_SEDs}). 
Finally, we summarize the model-dependent upper limits on $A_V$ for all the four targets (Section~\ref{subsec:comparison}).}

\subsection{Multi-wavelength SEDs of LRDs}
\label{subsec:SED_LRD}

\blue{We begin by the case for the brightest known LRD, A2744-45924, 
as our primary example to demonstrate the SED analysis.}
\blue{The results for the second individual target, RUBIES-BLAGN-1, are qualitatively similar and are presented 
in the summary analysis in Section~\ref{subsec:comparison}.}
In the left panel of Figure~\ref{fig:SED with Different Av}, we present both the incident SED and the corresponding IR re-emission spectra for our fiducial model, which adopts $(\gamma,n_0)=(0.5,10^3~\cc)$ and the SMC opacity law.
As the value of $A_V$ increases from $0.2$ to $1.0 ~ {\rm mag}$, the incident (de-reddened) SED inferred from the 
observed SED using Equation~(\ref{eqn:de-extinguish}) becomes brighter significantly at shorter wavelengths due to the steep UV rise in the SMC extinction curve (see Figure~\ref{fig:opacities}). 
The corresponding energy absorbed by dust is reprocessed and re-emitted in the IR band, shown by solid curves. 
For $A_V \le 0.4 ~ {\rm mag}$, the re-emitted IR flux remains below the observed flux data.
However, for $A_V\sim 0.6 ~{\rm mag}$, the predicted IR flux approaches or exceeds the MIRI F2100W detection, thereby violating the observational constraint. 
Using the method described in Section~\ref{subsec: A_V limit}, we determined the maximum allowed extinction for this dust distribution to be $A_V=0.55 ~ {\rm mag}$.

\begin{figure*}[t]
\centering
\includegraphics[width=0.49\textwidth]{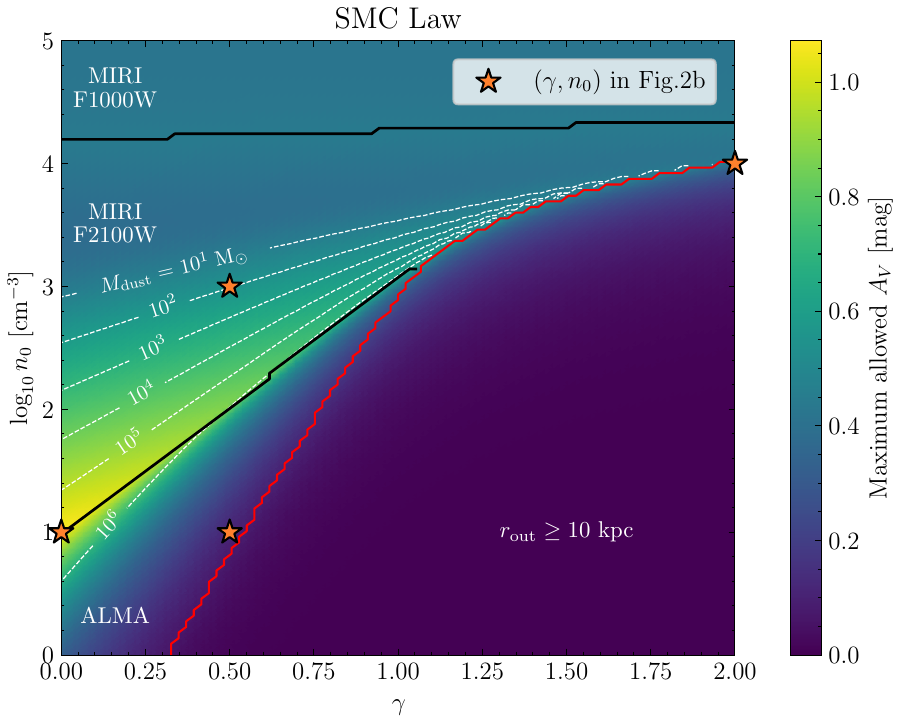}
\includegraphics[width=0.49\textwidth]{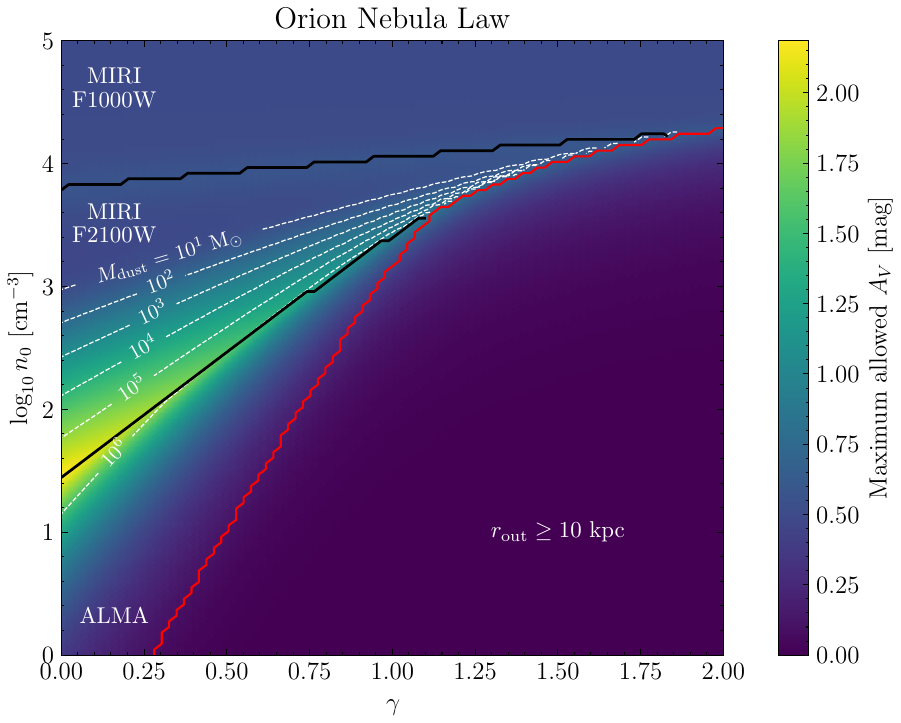}
\caption{Parameter survey showing the upper limit on $A_V$ as a function of the density distribution parameters $(\gamma, n_0)$,
assuming the SMC (left) and Orion Nebula (right) extinction curves for LRD A2744-45924.
Black solid curves divide the regions where the maximum allowed $A_V$ value is constrained by MIRI F1000W, F2100W, or ALMA data (from the top to the bottom).
White dashed contours label the total dust mass $M_{\rm dust}$, computed using Equation~(\ref{eq:Mdust}) based on the corresponding $A_V$ upper limit. 
Orange stars mark the density configurations shown in the right panel of Figure~\ref{fig:SED with Different Av}. 
The right-bottom region (i.e., steeper profiles with lower density normalization) requires extended dust distribution with $r_{\rm out}\geq 10$ kpc to achieve sufficient column density,
and is therefore ruled out as unphysical configurations. 
}
\label{fig:paras_survey}
\vspace{3mm}
\end{figure*}

In the right panel of Figure~\ref{fig:SED with Different Av}, we compare four representative models with different values of $(\gamma, n_0)$. 
For each case, its re-emission IR spectrum is given with the upper limit of $A_V$.
We present the same model as in the left panel where $(\gamma,n_0)=(0.5,10^3~\cc)$, but with the upper limit of $A_V = 0.55~{\rm mag}$. This parameter combination leads to a high average dust temperature of $T_{\rm dust}\gtrsim 200~\K$, resulting in strong emission in the NIR. 
\blue{The second model for $(\gamma, n_0) = (0.5, 10~\cc)$ with a lower $n_0$ value has a significantly extended and diffuse dust distribution for maintaining a given column density (and $A_V$).}
\blue{This leads to significant FIR re-emission from the outer region where the mass-weighted IR emissivity is higher}, tightly constrained by ALMA non-detection. Thus, a stricter upper limit of $A_V\simeq 0.11 ~{\rm mag}$ is given in this case. 
\blue{The third model for $(\gamma, n_0) = (0, 10~\cc)$ with a flatter profile 
shifts most energy into the MIR band}.
Since the Herschel constraints in this band are relatively shallow, the corresponding $A_V$ upper limit 
is more relaxed to $1.08 ~{\rm mag}$.
Yet, this value remains substantially lower than $\sim 3 ~{\rm mag}$ typically assumed in previous studies \citep{Setton_2025b}.
The fourth model assumes a dense and concentrated density profile with $(\gamma, n_0) = (2.0, 10^4~\cc)$. This model is motivated by the physical connection to dense nuclear scales (see Section~\ref{subsec:connect_to_inner_nucleus}). While this SED shows rest-frame NIR bump, the flux density remains consistent with JWST MIRI data (F1000W and F2100W) due to moderate extinction of $A_V\simeq 0.41~{\rm mag}$.


\subsection{Parameter Surveys}
\label{subsec:param_surveys}

Here, we systematically investigate how the upper limit on $A_V$ depends on the dust distribution parameters, $n_0$ and $\gamma$, considering both the SMC and Orion Nebula extinction laws. 
The observational constraints are taken from JWST/MIRI, Herschel/PACS, and ALMA data for A2744-45924 \citep{Labbe_2024_MostBrightLRD,Setton_2025b}. 
Figure~\ref{fig:paras_survey} summarizes the maximum allowed $A_V$ for each combination of $(\gamma, n_0)$.
The star symbols mark the parameter sets corresponding to the four models shown in the right panel of Figure~\ref{fig:SED with Different Av}.
We also overlay contours (white dashed) of constant dust mass, estimated from the density distribution parameters ($\gamma,n_0$) and the associated $A_V$ upper limit.

The left panel of Figure~\ref{fig:paras_survey} shows the $A_V$ upper limits derived using the SMC extinction law. 
The allowed values of $A_V$ vary significantly across the parameter space. 
In the upper-left region, characterized by high $n_0$ and low $\gamma$, the $A_V$ constraint is primarily given by MIRI observations and becomes tighter, as these models produce strong NIR/MIR emission from warm dust $T_{\rm dust}\gtrsim 200~\K$ and exceed the observed fluxes unless $A_V$ is sufficiently low.
In contrast, in the lower-right region, low $n_0$ and high $\gamma$, the constraint is dominated by ALMA non-detections, since these models yield significant FIR emission from colder dust ($T_{\rm dust}\gtrsim 30~\K$). 
The black solid line marks the boundary separating the MIRI- and ALMA-constrained regimes. 
The right-hand side of the red curve (i.e., steeper profiles with lower $n_0$) indicates the region where the dust distribution is far extended to $r_{\rm out}\geq 10$ kpc to achieve sufficient column density. We rule out the domain as unphysical configurations. 
For the SMC opacity model, the least restrictive constraint appears at $(\gamma, n_0) \approx (0, 10~\cc)$, where the upper limit reaches $\AVlim \simeq 1.07~{\rm mag}$.

The right panel of Figure~\ref{fig:paras_survey} presents the results assuming the Orion Nebula extinction law.
A key difference from the SMC case is that the Orion extinction law generally yields a higher $A_V$ upper limit, reaching up to $\simeq 2~{\rm mag}$ around $(\gamma, n_0) \approx (0, 30~\cc)$.
This difference is primarily because the Orion extinction curve is significantly flatter in the UV range (see Figure~\ref{fig:opacities}).
As a result, for a given $A_V$, the de-reddened intrinsic AGN spectrum is less luminous in the UV, leading to weaker reprocessed IR emission. 
Therefore, a higher $A_V$ value is allowed without reprocessed IR fluxes exceeding the observed fluxes.

In both cases, the upper limit of $A_V$ is obtained for a flat density profile ($\gamma\simeq 0$) with a normalization at $n_0\simeq 10-30~\cc$.
Given the dust distribution and corresponding $A_V$ limit, the total dust mass is calculated by integrating the density profile from $r_{\rm in}$ to $r_{\rm out}$ (see Equation~\ref{eq:Mdust}), as $M_{\rm dust}\sim 10^{5-6}~\msun$.
In the region below the boundary (\blue{red} solid curve), the required dust mass increases and reaches up to $M_{\rm dust}<10^7~\msun$\footnote{The right-bottom region requires substantially extended dusty distributions with $r_{\rm out}\geq 10$ kpc. Such large radii yield infinitely large dust mass because $M_{\rm dust}\propto r_{\rm out}^{3-\gamma}$ for $\gamma<3$.}. 
By contrast, the high-density regime allows for significantly lower dust mass.
That is because the outer edge of the dusty medium becomes smaller in order to achieve the required column density for extinction, thereby reducing the total dust mass.
In Section~\ref{subsec:dust_mass}, we use the dust mass estimates to place additional constraints on the upper limit on $A_V$.

We also note that previous studies have estimated dust masses for LRDs to be $M_{\rm dust} = 10^{3-5}~\msun$ \citep{Casey_2024_DustInLRD},
assuming visual extinction $A_V\sim 3-4~{\rm mag}$.
These estimates are based on an effective size of $\langle R_{\rm e}\rangle \simeq 100~\pc$ measured in the F444W filter. 
The dust mass derived in our study is broadly consistent with these previous results, but tends to be higher by a factor of $\sim 10$.
This is because our model determines the dust mass based on the physical extent of the dusty region, $r_{\rm out}$, which likely 
exceeds the effective size measured in the rest-frame optical, where dust does not emit. Recently, \cite{Casey_2025_DustMassUpper} shows that the dust mass in LRDs is limited to $< 10^6~\msun$ based on the non-detection of stacking deep ALMA observations for 60 LRDs, aligning well with our analysis.

\subsection{\blue{$A_V$ upper limits for the SEDs of stacked LRDs}}
\label{subsec:stacked_SEDs}

\begin{figure*}[t]
\centering
\includegraphics[width=0.51\textwidth]{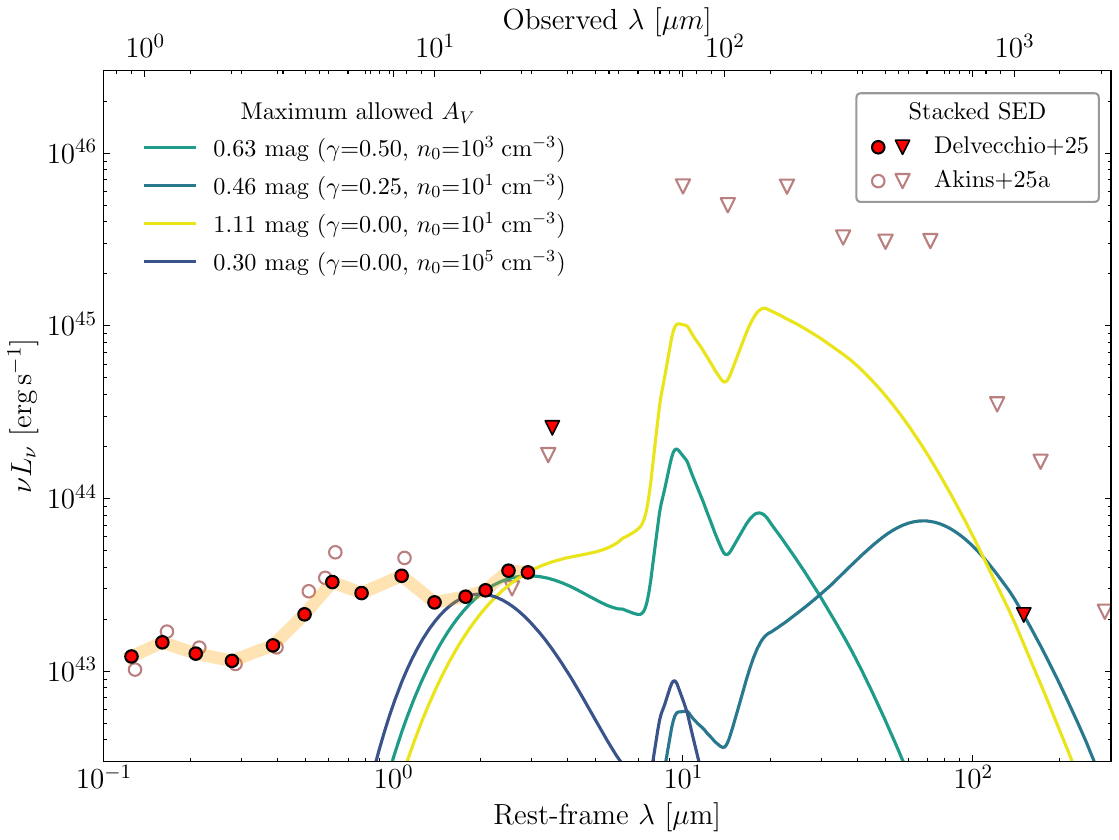}
\includegraphics[width=0.47\textwidth]{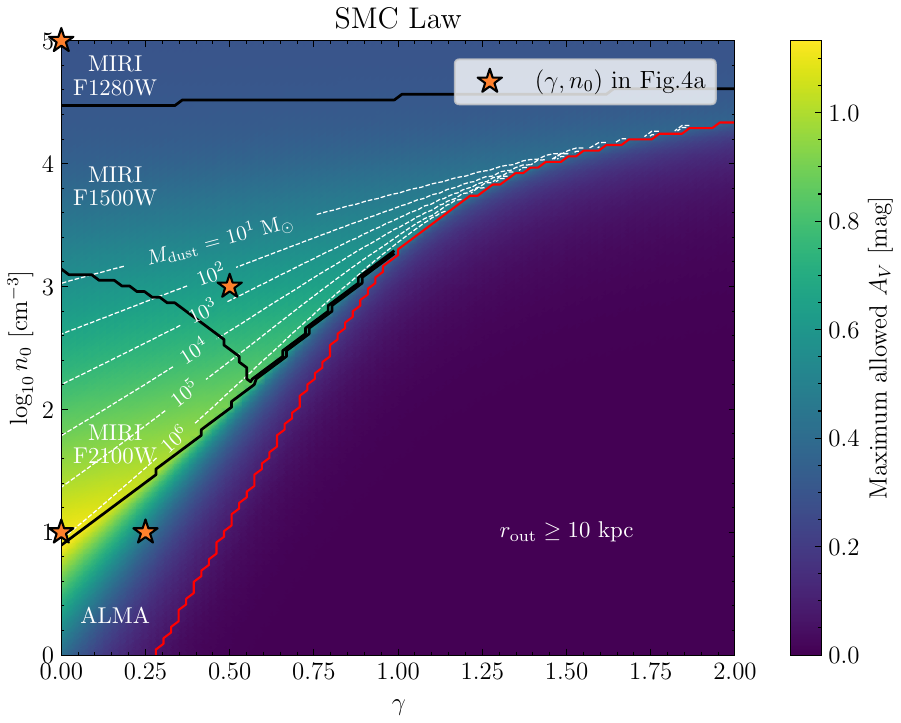}
\caption{
\blue{
    {\it Left}: Infrared re-emission SEDs with the maximum allowed $A_V$ for different dust density configurations, constrained with the stacked LRD SED from \cite{Delvecchio_agn-heated_2025} (filled circles and triangles). 
    Stacked SED from \cite{Akins_COSMOS-Web_2025} is also shown for comparison (open circles and triangles).  
    The parameter combinations are chosen as $(\gamma, n_0)=(0.5,10^3~\cc)$, $(0.25,10~\cc)$, $(0,10~\cc)$, and $(0, 10^5~\cc)$. 
    {\it Right}: Parameter survey showing the upper limit on $A_V$ as a function of the density distribution parameters $(\gamma, n_0)$, assuming the stacked LRD SED and the SMC opacity curve. 
    Black solid curves divide the regions where the maximum allowed $A_V$ value is constrained by MIRI F1280W, F1500W, F2100W, or ALMA data (from the top to the bottom). \\
}
}
\label{fig:stacking}
\vspace{3mm}
\end{figure*}

\blue{Next, we extend our analysis to the stacked SEDs and discuss whether the stringent upper limit of dust contents is generally applied to broader LRD population.
The left panel of Figure~\ref{fig:stacking} presents the stacked SED (circles for detections and triangles for non-detections; data taken from \citealt{Delvecchio_agn-heated_2025} and \citealt{Akins_COSMOS-Web_2025}),
together with the SED models computed for four different parameter combinations of ($\gamma,n_0$), which are constrained using the data from \citet{Delvecchio_agn-heated_2025}.
As in the cases shown in the right panel of Figure~\ref{fig:SED with Different Av} (see also Section~\ref{subsec:SED_LRD}), the model with a flat and extended dust distribution $(\gamma \simeq 0, n_0 \simeq 10~\cc)$ yields the least restrictive (i.e., largest) upper limit on $A_V$.
Such dust density profiles produce MIR spectra with a broader peak 
that delicately balances the constraints from both MIRI detections and ALMA non-detections, thereby allowing a higher peak of reprocessed energy.}


\blue{It is worth noting the difference between the two stacked SEDs and how the results depend on the choice of dataset.
The SED dataset of \citet{Delvecchio_agn-heated_2025} covers the UV-optical-NIR wavelength ranges, thank to the inclusion of additional MIRI photometry, which is overall consistent with that of \citet{Akins_COSMOS-Web_2025} but includes a larger number of multi-band data points.
Beyond the rest-frame wavelength of $\sim 3~\mum$, even the stacked SEDs provide only upper limits due to non-detections of LRDs, although \citet{Delvecchio_agn-heated_2025} added a more stringent upper limit from MIRI/F2100W.
In the mid- to far-infrared regime, where non-detection limits from Spitzer, Herschel, and SCUBA-2 are available \citep{Akins_COSMOS-Web_2025}, these limits are too shallow to provide any additional constraints beyond those already imposed by the MIRI detections.
This highlights the critical need for next-generation far-infrared observatories with improved sensitivity to robustly constrain the dust content in these enigmatic objects, a point we will revisit in Section~\ref{subsec:PRIMA_test}.}


\blue{The right panel of Figure~\ref{fig:stacking} shows the full parameter survey constrained with data from \cite{Delvecchio_agn-heated_2025} and SMC opacity, where the colored markers correspond to the models plotted in the left panel and black lines demarcate regions constrained by different observational data. 
Overall, the maximum allowed visual extinction is tightly constrained to $A_V \lesssim 1.13~{\rm mag}$.
We performed an identical analysis on the stacked SED from \citet{Akins_COSMOS-Web_2025} (not shown for brevity). 
This yielded a similarly stringent but slightly more relaxed upper limit of $A_V \lesssim 1.22~{\rm mag}$. 
These results are broadly consistent with those obtained for A2744-45924 and RUBIES-BLAGN-1 in Section~\ref{subsec:param_surveys}, suggesting that the low dust content inferred for the two individual LRDs may be representative of the overall LRD population.}

\subsection{Model-dependent $A_V$ Upper Limits}
\label{subsec:comparison}

\begin{figure}
    \centering
    \includegraphics[width=85mm]{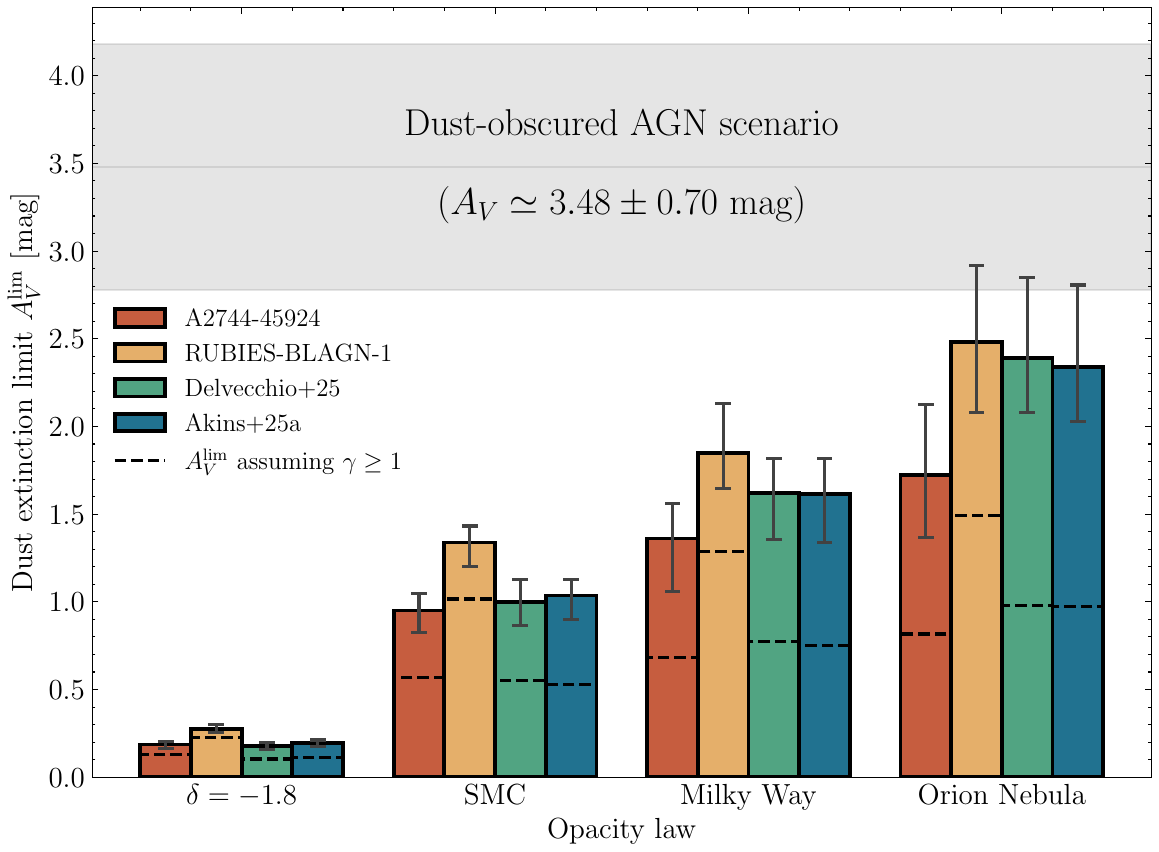}
    \caption{
    \blue{
    Summary of the upper limits on $A_V$ derived for the four different SEDs analyzed in this work: the two individual bright LRDs (A2744-45924 and RUBIES-BLAGN-1) and the two stacked LRD samples from \citet{Delvecchio_agn-heated_2025} and \citet{Akins_COSMOS-Web_2025}.
    The bar chart compares the results across the four dust opacity laws (from the left to the right: a modified Calzetti law with $\delta = -1.8$, SMC, Milky Way, and Orion Nebula). 
    Each solid rectangle shows the least restrictive $A_V$ upper limit ($\AVlim$) found from the entire $(\gamma, n_0)$ parameter space with an error bar representing $\AVlim$ under different dust mass limits: $M_{\rm dust} \leq 10^4$, $10^5$, and $10^6~\msun$ from the bottom to the top. 
    The dashed lines indicate the more stringent limit obtained when restricting the analysis to physically-motivated, concentrated density profiles ($\gamma \geq 1$).
    }
    The gray shaded region indicates the extinction range typically assumed in obscured AGN scenarios ($A_V \simeq 3.48\pm 0.70~{\rm mag}$, \citealt{Akins_COSMOS-Web_2025}). 
    }
    \label{fig:summary}
\end{figure}

\blue{
Figure~\ref{fig:summary} summarizes the upper limits on $A_V$ for all four targets analyzed in this work: the two individual LRDs (A2744-45924 and RUBIES-BLAGN-1) and the two stacked SEDs from \citet{Delvecchio_agn-heated_2025} and \citet{Akins_COSMOS-Web_2025}. 
Four different dust extinction laws are considered (from the left to the right: a modified Calzetti law with $\delta = -1.8$, SMC, Milky Way, and Orion Nebula), as introduced in Section~\ref{subsec:dust_opacity}. 
For each case, we show the least restrictive $A_V$ upper limits across the entire $(\gamma, n_0)$ parameter space, represented by colored bars. 
The error bars indicate $\AVlim$ values under the constraint of total dust mass, $M_{\rm dust} \leq 10^4$, $10^5$, or $10^6~\msun$ from the bottom to the top. 
As expected, tighter constraints on $M_{\rm dust}$ result in a more stringent limit on the allowed $A_V$ values. 
}

For A2744-45924, the $A_V$ upper limit increases from $\simeq 0.95^{+0.10}_{-0.13}~{\rm mag}$ for the SMC law to $\simeq 1.81^{+0.31}_{-0.37}~{\rm mag}$ for the Orion Nebula law. 
As described in Section~\ref{subsec:param_surveys}, this trend originates from the differences in the UV slope of the extinction curve characterized by the ratio of $A_{\rm UV}/A_V$, where $A_{\rm UV}$ is the extinction level at $1215~{\rm \AA}$. 
The SMC curve has a steep UV slope with $A_{\rm UV}/A_V \simeq 6.8$, while the MW and Orion curves are flatter ($\simeq 3.4$ and $\simeq 1.8$, respectively).
Therefore, a steeper extinction curve in UV (i.e., higher $A_{\rm UV}/A_V$) results in stronger UV attenuation for a given $A_V$, 
leading to more reprocessed IR fluxes and thus to a stricter $A_V$ upper limit.
The extinction model with $\delta=-1.8$ of Equation~(\ref{eq:noll}) corresponds to a case with an extremely large slope of the extinction curve, $A_{\rm UV}/A_V\simeq 44$.
Our analysis requires $A_V \leq 0.2~{\rm mag}$; otherwise, the reprocessed IR emission exceeds the observational constraints.

For RUBIES-BLAGN-1, the $A_V$ upper limits are generally higher than those for A2744-45924, ranging from 
$\simeq 1.34^{+0.11}_{-0.13}~{\rm mag}$ (SMC) to $\simeq 2.58^{+0.27}_{-0.36}~{\rm mag}$ (Orion). 
This results from the fact that RUBIES-BLAGN-1 exhibits weaker observed fluxes in rest-frame UV,
which leads to a lower inferred intrinsic radiation flux for a given $A_V$.
As a result, higher $A_V$ values are permitted without exceeding the observed IR constraints.

\blue{Our analysis of the two stacked SEDs from \citet{Delvecchio_agn-heated_2025} and \citet{Akins_COSMOS-Web_2025} yields comparable upper limits on $A_V$, as shown in Figure~\ref{fig:summary}.
For instance, the limits for the \citet{Delvecchio_agn-heated_2025} sample range from $\AVlim \simeq 1.13^{+0.13}_{-0.13}~{\rm mag}$ (SMC) to $\simeq 2.39_{-0.31}^{+0.46}~{\rm mag}$ (Orion).
The broad consistency between the limits derived from these large-sample averages and those from the individual sources provides strong evidence that our findings generalize beyond specific examples and are representative of the broader LRD population.}

We also remind readers that the least restrictive $A_V$ is obtained for flat density profiles ($\gamma \simeq 0$) with $n_0 \simeq 10-30~\cc$. 
However, such profiles might be unphysical configurations, especially when considering the physical connection to the inner nucleus, 
see discussion in Section~\ref{subsec:connect_to_inner_nucleus}. 
Therefore, we also show the results for more concentrated density profiles with $\gamma \geq 1$ \blue{(dashed horizontal lines)}. 
These results are more stringent, yielding $\sim 50\%$ lower $A_V$ upper limits.

As examined so far, the $A_V$ upper limit depends on dust extinction laws, dust density distributions, dust mass requirements, and individual spectral features of LRDs.
Nevertheless, the overall trend robustly suggests that the upper limit $\AVlim \simeq 1~{\rm mag}$ in our analysis 
is significantly lower than the average value of $\langle A_V\rangle \simeq 3.48~{\rm mag}$ obtained from $\sim 500$ photometrically-selected LRD samples based on SED fitting analysis assuming obscured AGN scenarios \citep[gray shaded region,][]{Akins_COSMOS-Web_2025}. 
In particular, with the SMC extinction law, the upper limit on $A_V$ is reduced by $\Delta A_V \equiv \langle A_V\rangle - \AVlim\simeq 2.5~{\rm mag}$.
We will discuss the implications of this result to energetics for the entire LRD population in Section~\ref{subsec:Soltan_argument}.

\section{Discussion}

\subsection{Where were dust grains from?}
\label{subsec:dust_mass}

Our SED model provides not only the reprocessed IR spectrum but also an estimate of the dust mass for a given dust distribution.
For the $A_V$ upper limits, the corresponding dust mass is typically on the order of $M_{\rm dust}\lesssim 10^{5-6}~\msun$.
Where do the amount of dust grains come from?
In galaxies, dust grains are considered to form through stellar winds of evolved stars and from supernova ejecta,
the latter of which is dominant in the early universe ($\lesssim 1~{\rm Gyr}$).
The dust production efficiency per unit stellar mass depends sensitively on the star formation history and the type of dust producers.
Theoretical models estimate a dust-to-stellar mass ratio in the range of $f_{\rm dust,\star}\simeq 10^{-5}-10^{-3}$, depending on whether supernova reverse shocks
effectively destroy newly created dust grain in ejecta \citep[see Figure~18 in][and reference therein]{Schneider_2024}. 
Assuming a fiducial value of $f_{\rm dust,\star} = 10^{-4}$ for a constant star formation rate, the stellar mass required to produce $M_{\rm dust}$ is given by
\begin{equation}
M_\star \simeq 10^{9}~\msun \left(\frac{M_{\rm dust}}{10^{5}~\msun}\right)
\left(\frac{f_{\rm dust,\star}}{10^{-4}}\right)^{-1}.
\end{equation}

For A2744-45924, there is no clear host galaxy signature, but a faint, extended component is observed offset from the central LRD. 
Spectral and morphological analysis indicates that its unresolved host galaxy, if present, has a stellar mass of $M_\star<10^{9.6}~\msun$ 
within an effective radius of $R_{\rm e}\simeq 70~\pc$ \citep{Labbe_2024_MostBrightLRD,Chen_2024_HostG_LRD}.
On the other hand, the extended component may be either a young low-mass galaxy ($M_\star \simeq 1.5\times 10^{8}~\msun$ 
and $t_{\rm age}\simeq 15$ Myr, if assuming the same redshift as in the LRD; otherwise, it is a foreground galaxy at $z_{\rm photo}\simeq 1.02$) 
or low-metallicity nebular emission ($Z\simeq 0.05~\zsun$) powered by the LRD \citep{Chen_nebula_2025}.
These interpretations imply that the associated dust mass should be relatively low, 
$M_{\rm dust} \lesssim (0.15$--$4)\times 10^5~\msun$.
For RUBIES-BLAGN-1, no host galaxy features are detected in imaging.
However, spectral fitting using AGN+stellar composite models suggests $M_\star \simeq 1.8\times 10^9~\msun$,
which is responsible for the observed rest-frame UV emission \citep{Wang_2024_z3LRDwMIRI}.
The stellar mass similarly implies a dust mass of $M_{\rm dust} \lesssim 1.8\times 10^5~\msun$.
In both cases, these dust mass estimates disfavor the least restrictive $A_V$ upper limit obtained for a
specific dust density profiles (see Figure~\ref{fig:paras_survey}), corresponding to $M_{\rm dust}\sim 10^6~\msun$. \citet{Akins_2025_tentative_ALMACI} provided a dust mass upper limit for A2744-45924 and RUBIES-BLAGN-1 as $M_{\rm dust} < 7\times 10^7~\msun$ and $<1.4\times10^8~\msun$ using the Rayleigh-Jeans tail
of the SED in ALMA bands, respectively, and reported a tentative detection of the narrow [\ion{C}{1}] line (809.34~GHz in rest-frame) with ${\rm FWHM}\simeq 80~\kms$ in A2744-45924, implying a dynamical mass of $\lesssim 10^{10}~\msun$.
These mass estimates are consistent with our analysis.
Future improvements of stellar mass measurements \blue{(e.g. deeper MIRI observations)} may provide more stringent limits on the dust content and extinction in LRDs,
making the dust budget crisis severer.

While our argument does not provide a lower limit on $A_V$ since we make no assumption about the intrinsic SED 
and instead reconstruct it through spectral de-reddening, some SED models for LRDs with a prominent Balmer break 
require extremely high hydrogen column densities $N_{\rm H}\gtrsim 10^{25}~{\rm cm}^{-2}$ but adopt only modest 
extinction levels of $A_V\sim 0.2\text{--}0.5~{\rm mag}$ \citep{Naidu_2025,Taylor_2025}. 
These models implicitly assume that the dust-to-gas mass ratio in such LRDs is $\lesssim 10^{-4}$ of 
the canonical solar-neighborhood value $\mathcal{D}_\odot =0.01$. 
This suggests that at least some LRDs might emerge from extremely low-metallicity (or pristine) environments 
with $Z\sim 10^{-4}~\zsun$ or under peculiar conditions where dust grains formation is strongly suppressed.
Recently, several theoretical studies have proposed spectral models that yield {\it intrinsically red} optical continua without relying on dust reddening:
dense gaseous envelopes surrounding BHs \citep{Kido_2025}, circum-binary disks around coalescing BHs \citep{Inayoshi_2025_BBH}, and self-gravitating disks \citep{Zhang_2025}.

\begin{figure*}
    \centering
    \includegraphics[width=140mm]{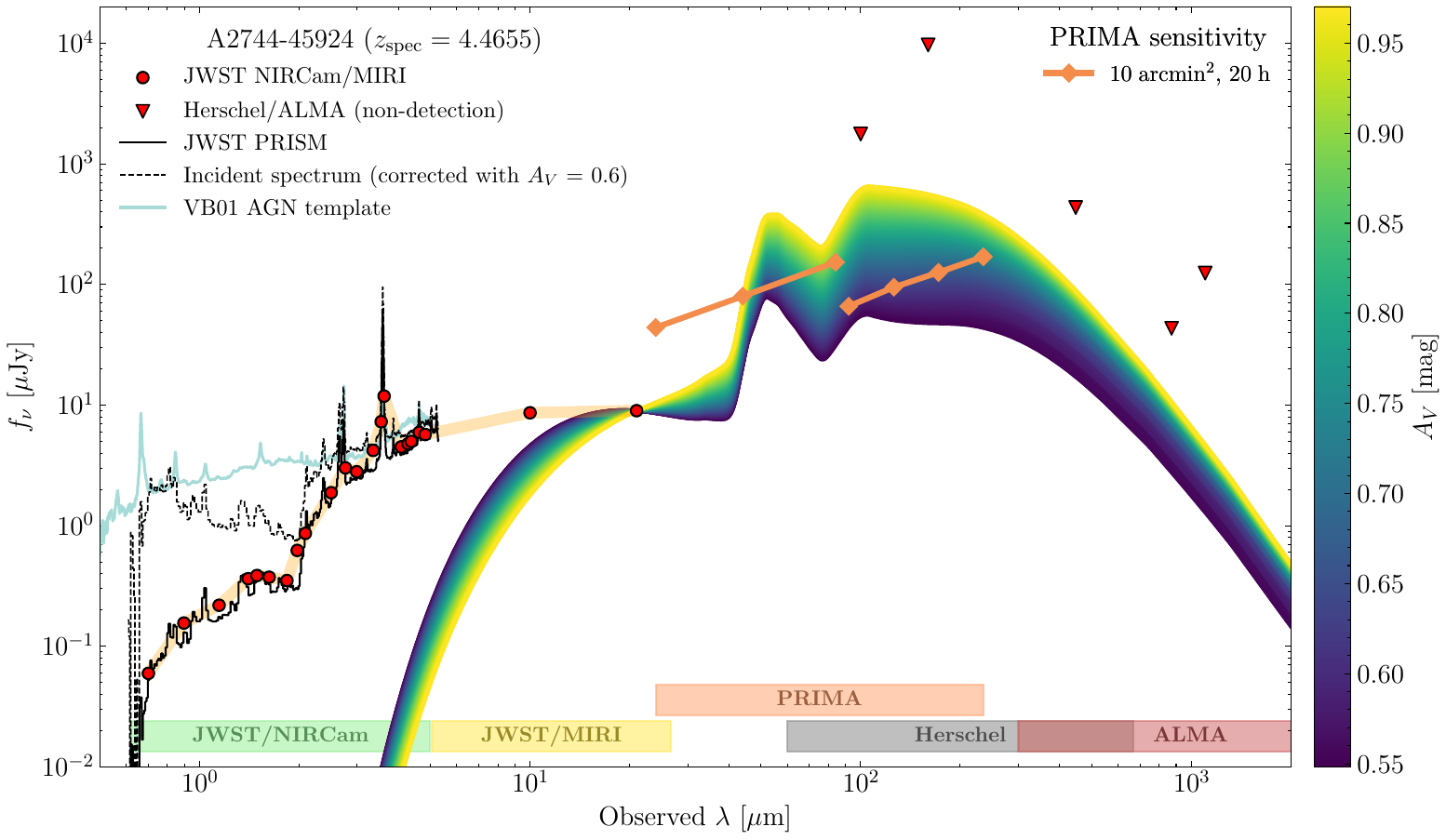}
    \caption{Observed flux densities ($f_\nu$) of the brightest LRD (A2744-45924) at $z_{\rm spec}=4.4655$, 
    comparing modeled dust re-emission with current observational data and future detection capabilities with PRIMA.
    Observed photometric data points for A2744-45924 are shown in circles and triangles, together with the JWST spectrum (black line). 
    The colored curves represent the maximum predicted IR re-emission flux densities from various dust distribution models, 
    each calculated to be consistent with existing observational constraints from JWST MIRI and ALMA, with the colors indicating the derived 
    upper limit on visual extinction ($A_V$) for each model.
    All models presented correspond to a fixed dust mass of $M_{\rm dust} = 10^5~\msun$.
    The black dashed curve presents the expected intrinsic spectrum with dust correction assuming $A_V=0.6~{\rm mag}$, which does not agree with the shape of the low-$z$ AGN composite spectral template \citep[cyan,][]{VandenBerk_2001}.
    The orange lines indicate the anticipated $5\sigma$ sensitivity for a 20-hour PRIMA exposure for $10~{\rm arcmin}^2$ areas.
    Colored boxes at the bottom illustrate the wavelength coverage of relevant telescopes and instruments.
    }
    \vspace{5mm}
    \label{fig:summary_SED}
\end{figure*}

\subsection{So{\l}tan--Paczy\'{n}ski Argument}
\label{subsec:Soltan_argument}

Compiling data from extensive AGN surveys, we constrain the radiative efficiency of BHs by comparing the observed BH mass density $\rho_{\rm BH}$ to the mass accretion rate to BHs inferred from the luminosity density $\mathcal{L}$ per comoving volume in earlier epochs, known as the So{\l}tan--Paczy\'nski argument \citep{Soltan_1982,Yu_Tremaine_2002}.
This analysis suggests a radiative efficiency $\epsilon_{\rm rad}\simeq 10\%$ for AGNs at $0<z<5$ and favors modest spins of these BHs under geometrically-thin disk configuration \citep{Novikov_Thorne_1973,Shakura_Sunyaev_1973}.
The same argument has been applied to LRD populations \citep{Inayoshi_Ichikawa_2024} by quantifying $\rho_{\rm BH}$ and $\mathcal{L}$ through integrating the LRD luminosity function and BH mass functions, respectively, under the assumption of obscured AGNs \citep{Matthee_2024,Kokorev_2024_LRDcensus,Akins_COSMOS-Web_2025,Greene_2024}.
The radiative efficiency results in $\epsilon_{\rm rad}\simeq 20\%$, suggesting rapid BH spins with 96\% of the maximum limit.

We here revisit the So{\l}tan-Paczy\'nski argument for LRD populations by accounting for our conclusion that the LRD spectral features do not necessarily require heavy dust reddening of AGNs.
Instead, the level of extinction must be modest ($A_V\lesssim 1~{\rm mag}$) to remain consistent with MIRI detection and ALMA non-detection.
This revised extinction enables a re-evaluation of both the intrinsic AGN luminosity and single-epoch virial BH mass, $M_{\rm BH} \propto L^{1/2}$ \citep[e.g.,][]{Kaspi_2005}, 
using the broad-line region luminosity-size empirical relation calibrated via reverberation mapping for nearby AGNs.
Here, $L$ refers to the AGN continuum luminosity at rest-frame 5100~{\AA} or the luminosity of the broad H$\alpha$ (H$\beta$) emission line \citep[e.g.,][]{Greene_Ho_2005,Reines_2013}, both of which are dust corrected using the inferred visual extinction $A_V$ as they lie at similar wavelengths. 

\blue{When the $A_V$ value is reduced by $\Delta A_V = \langle A_V\rangle - \AVlim$ (see Section~\ref{subsec:comparison}), the intrinsic AGN luminosity would decrease by a factor of $10^{0.4 \Delta A_V}$, leading to a decrease in the estimated BH mass by a factor of $10^{0.2 \Delta A_V}$. 
In particular, assuming SMC extinction law for A2744-45924, the upper limit on $A_V$ is reduced by $\Delta A_V \simeq 2.5~{\rm mag}$, so the estimated BH mass would 
decrease by a factor of $\sim 10^{0.5} \approx 3.16$.}

Therefore, the impact on the inferred radiative efficiency $\epsilon_{\mathrm{rad}}$ can be quantified as
\begin{align}
\frac{\epsilon_{\rm rad}}{1-\epsilon_{\rm rad}} &= \frac{\int_0^{t_{0}} \mathcal{L} (t) dt}{\rho_{\rm BH}(t_0)c^2}
\simeq 0.25 \cdot 10^{-0.2\Delta A_V},\nonumber\\[5pt]
& \rightarrow  0.079~~~~{\rm with}~\Delta A_V=2.5~{\rm mag},
\end{align}
where $t_0$ is the cosmic time at $z=4$ when the BH mass density is measured, and the prefactor of $0.25$ is the best-fitted value \citep[$\epsilon_{\rm rad}\simeq 20\%$;][]{Inayoshi_Ichikawa_2024} under the assumption of obscured AGNs with $\langle A_V\rangle \simeq 3.48~{\rm mag}$ \citep{Akins_COSMOS-Web_2025}.
With the revised $A_V$ constraint in this work, we find a radiative efficiency of $\epsilon_{\rm rad}\simeq 0.073$, which aligns well with 
the canonical 10\% radiative efficiency \citep{Yu_Tremaine_2002,Ueda_2014,Delvecchio_2014}\footnote{This low value of $\epsilon_{\rm rad}\simeq 0.073$ is consistent with that of non-spinning BHs under the thin-disk approximation, or possibly indicates a low radiative efficiency due to photon trapping in super-Eddington slim disks around BHs with moderate spins of $a_{\rm BH}\simeq 0.7$ \citep{Abramowicz_1988,Ohsuga_2005,Inayoshi_ARAA_2020}}.
Thus, this resolves the apparent tension with the classical So{\l}tan argument for the LRD population \citep{Inayoshi_Ichikawa_2024}.

\subsection{Physical Connection to the Inner Nucleus}
\label{subsec:connect_to_inner_nucleus}

We examine how the $A_V$ upper limit depends on the dust density distribution, using the observed spectral of
A2744-45924 and RUBIES-BLAGN-1, both of which show a Balmer break feature \citep{Wang_2024_z3LRDwMIRI,Labbe_2024_MostBrightLRD}.
A recently proposed interpretation attributes this spectral break to absorption by dense gas with $n_{\rm cl}\gtrsim 10^{9}~\cc$,
where atomic hydrogen is excited to the $n=2$ states via particle collision \citep{Inayoshi_Maiolino_2025,Ji_2025_BlackThunder}.
These absorbers can imprint the Balmer-break feature on their spectra as well as Balmer absorption on H$\beta$ and H$\alpha$ emission.
This scenario motivates us to explore the possible physical connection between the dense gas clouds in or just outside the broad-line region
(BLR; $r_{\rm BLR}\sim 0.01~\pc$) and the dusty media near the sublimation radius ($r_{\rm sub}\sim 1~\pc$).
Assuming that the gas density follows a power-law radial profile with the same index $\gamma$ as in the dusty region, the density at $r=r_{\rm sub}$ is given by
\begin{equation}
n_0 \simeq \frac{n_{\rm cl}}{\mathcal{C}}\left(\frac{r_{\rm BLR}}{r_{\rm sub}}\right)^\gamma,
\end{equation}
where $\mathcal{C}$ is the density enhanced factor caused by cloud clumpiness and shock-compression in outflowing materials.
Here, we set $\mathcal{C}=10$, though the true value may be smaller given the high covering fraction required by the absorption features \citep{Maiolino_2025_xrayJWSTAGN,Inayoshi_Maiolino_2025}.

With this scaling, a profile slope of $\gamma = 2$ suggests $n_0 \simeq 10^4~\cc$, while shallower slopes ($\gamma < 2$) yield even higher $n_0$ values. 
As shown in Figure~\ref{fig:paras_survey}, the least constrained $A_V$ upper limits arise from a flat density profile ($\gamma \simeq 0$) with $n_0 \simeq 10\text{--}30~\cc$.
However, if the density profile indeed connects the BLR and dust sublimation scales as suggested above, such flat profiles are disfavored.
In particular, for $\gamma = 2$ and $n_0 = 10^4~\cc$, the extinction level required to satisfy the MIRI and ALMA constraints is significantly lower, 
with $\AVlim \simeq 0.41~{\rm mag}$.

\subsection{\blue{LRD Models without Heavy Dust Extinction}}

\blue{The initial hypothesis for the red colors of LRDs invoked significant dust extinction ($A_V\sim 3$; e.g., \citealt{Kocevski_2023,Matthee_2024}),
but this interpretation is challenged by the weakness of rest-frame near- to mid-IR emission constrained with JWST/MIRI observations
\citep[e.g.,][]{Perez-Gonzalez_2024_MIRILRD,Williams_2024,Akins_COSMOS-Web_2025,Casey_2025_DustMassUpper,LiZ_2025_LRDDustSED}.
As an alternative, \cite{Inayoshi_Maiolino_2025} proposed that dense gas surrounding the AGN can imprint a prominent Balmer break 
when the absorber density reaches $n\sim 10^{9-10}~\cc$, largely independent of dust or metal content.
Following work \citep{Inayoshi_2025_BBH,Kido_2025} extended this framework to the optically thick regime, where hydrogen opacity (especially via H$^-$ ions)
sets the photospheric temperature to $T_{\rm eff}\simeq 5000~\K$, analogous to the Hayashi line known for protostars and post-main-sequence stars \citep{Hayashi_1961}.
Almost concurrently, several observational groups reported LRDs with extremely deep Balmer breaks (e.g., \citealt{Naidu_2025,deGraaff_2025,Taylor_2025}), 
inconsistent with evolved stellar populations but well explained by dense-gas models with column densities of $N_{\rm H}\sim 10^{25-26}~{\rm cm^{-2}}$.
In this interpretation, spectral reddening arises from hydrogen Rayleigh scattering (private communication), although one-dimensional plane-parallel CLOUDY calculations
are not strictly valid when scattering dominates.}

\blue{
While both the heavy dust extinction and dense gas envelope scenarios can redden the spectra of LRDs by absorbing and scattering UV/optical photons,
the two scenarios differ fundamentally in how the absorbed energy is reprocessed.
Dust grains thermalize the radiation and re-emit it in the infrared, which is tightly constrained by JWST/MIRI, Herschel, and ALMA observations. 
In contrast, a dense gaseous envelope primarily channels the absorbed energy into emission lines, thereby avoiding the IR continuum limits.
Using this energy balance argument, we constrain the dust contribution in LRDs to $A_V \lesssim 1.0-1.5~{\rm mag}$, significantly lower than previously assumed values. 
This finding implies that dust alone cannot account for the observed red slopes without violating IR limits, and that a dense, nearly dust-free envelope 
is the more plausible explanation for the spectral reddening.
}

\subsection{Direct Test with PRIMA}
\label{subsec:PRIMA_test}

Figure~\ref{fig:summary_SED} presents the compiled SED models for A2744-45924 at $z_{\rm spec}=4.4655$ that are consistent with 
the observational constraints from the JWST MIRI and ALMA.
The color code indicates the upper limits on $A_V$ for a fixed dust mass of $M_{\rm dust}=10^5~\msun$, assuming the SMC extinction curve.
For illustration, the expected intrinsic spectrum with dust correction assuming $A_V = 0.6~{\rm mag}$ is shown (black dashed), which does not match the low-$z$ AGN spectrum template \citep[cyan,][]{VandenBerk_2001}. 

Despite the deep MIRI and ALMA observations, the rest-frame MIR spectral shape remains largely unconstrained
due to the relatively shallow limits from Herschel data.
The PRobe for-Infrared Mission for Astrophysics (PRIMA) is expected to significantly improve MIR sensitivity 
for extragalactic astronomy\footnote{https://prima.ipac.caltech.edu/}.
We overlay the expected $5\sigma$ sensitivity of a 20-hour survey over $10~{\rm arcmin}^2$ using the PRIMAger mode, 
which offers hyperspectral narrow-band imaging ($R=10$) from 24 to 84~$\mum$ and 
four polarimetry broad band channels between $96$ and $235~\mum$ \citep[see more details in][]{PRIMA_2023}\footnote{https://prima.ipac.caltech.edu/page/etc-calc}. 
The PRIMA detection limit of $\sim 100~\muJy$ is more than two orders of magnitude deeper than that 
of Herschel/PACS under a similar survey design.
This PRIMA capability enables MIR detection for most of the SED cases presented, bringing opportunities 
to investigate key diagnostic features such as the $9.7~\mum$ silicate emission/absorption and PAH emission
(although our current SEDs do not have PAH features).

If the MIR fluxes remain undetected by the PRIMA survey with the expected detection threshold, 
the upper limit on $A_V$ will be further tightly constrained to $A_V\lesssim 0.62~{\rm mag}$.
The future PRIMAger survey program is an ideal platform to test our model and related theoretical interpretations for the LRD properties.

Furthermore, the modeling framework presented herein is both robust and readily applicable, offering a powerful tool to derive crucial upper limits on dust extinction and dust mass for any LRD with UV-to-IR photometric or spectroscopic data available. 
Such constraints are vital for unveiling the true nature of LRDs and will be valuable for maximizing the scientific yield from ongoing and upcoming surveys.

\begin{acknowledgments}
We greatly thank Hollis B.~Akins, Changhao Chen, Kohei Ichikawa, Yue Liu, Masafusa Onoue, and Jinyi Shangguan for constructive discussions. 
\blue{We also thank the anonymous referee for insightful comments that helped to improve the Letter.} 
K.~I. and L.~C.~H. acknowledge support from the National Natural Science Foundation of China (12233001), 
the National Key R\&D Program of China (2022YFF0503401), and the China Manned Space Program (CMS-CSST-2025-A09).
This work is supported by High-performance Computing Platform of Peking University.
The modeling code developed in this work is publicly available at \url{https://github.com/c-kj/LRD_dust_modeling} \citep{chen_2025_17380460}.
\end{acknowledgments}

\begin{contribution}
The first two authors (K.~Chen and Z.~Li) equally contributed to this work. 
\end{contribution}

\software{Astropy \citep{astropy:2013,astropy:2018,astropy:2022}, dust\_extinction\citep{Gordon_2024_package}, 
Cloudy \citep{CLOUDY_2023}, 
NumPy \citep{harris2020array}, 
SciPy \citep{2020SciPy-NMeth}, 
Matplotlib \citep{Hunter:2007}, 
LRD\_dust\_modeling \citep{chen_2025_17380460}.}

\appendix
\restartappendixnumbering

\section{Examining Alternative Scenario for UV Radiation Origin}

\blue{We here examine alternative cases where the observed UV flux have different origin from dust-obscured AGNs. 
The UV radiation could be contributed by either the AGN (through scattering or leakage) or the host galaxy, or a combination of both \citep{Akins_COSMOS-Web_2025,Kocevski_2023,Kokorev_2024_LRDcensus}. 
In both the scenarios, the intrinsic AGN UV flux is lower than the dust-dereddened flux from the observed value (Equation~\ref{eqn:de-extinguish}), 
allowing for a higher limit for $A_V$.}

\begin{figure*}[h]
    \centering
    \includegraphics[width=0.49\textwidth]{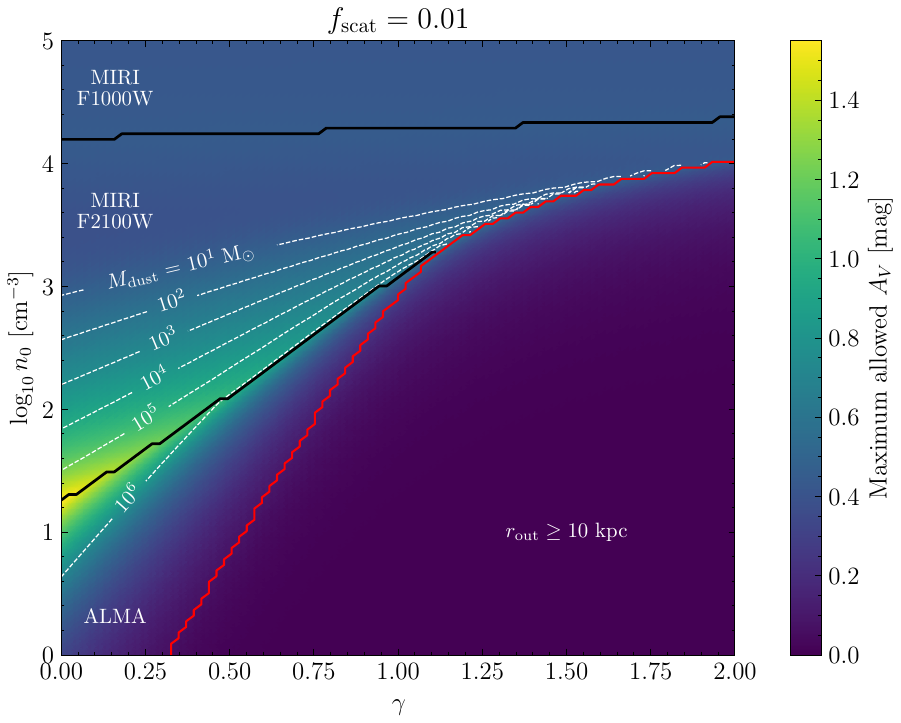}
    \includegraphics[width=0.49\textwidth]{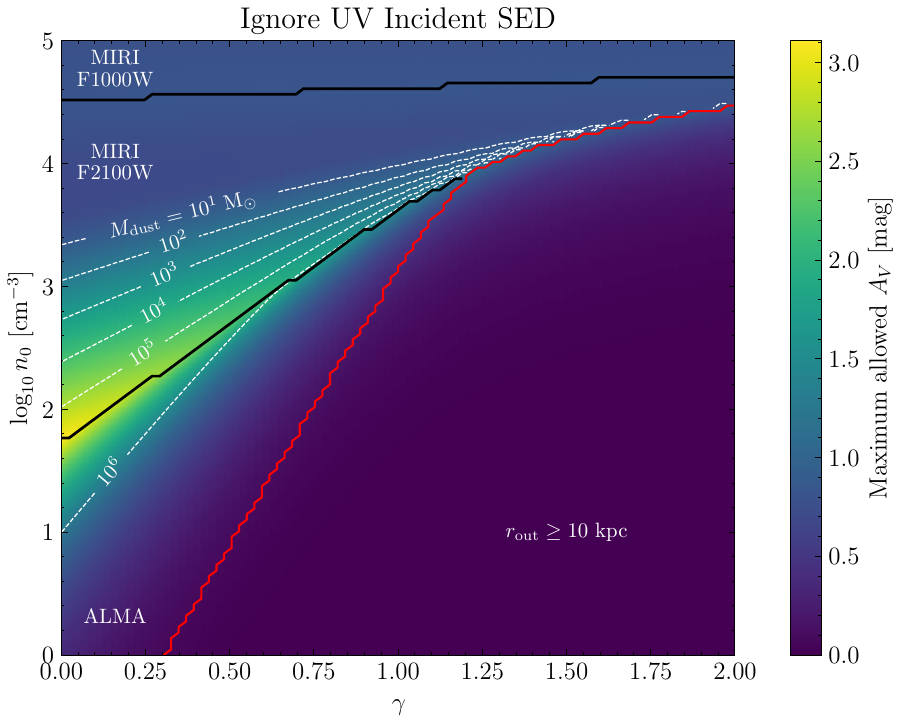}
    \caption{
        \blue{
            Similar to the left panel of Figure~\ref{fig:paras_survey}, but examining two alternative treatments of the intrinsic UV flux that is reprocessed to IR. 
            {\it Left}: Assuming a scattering fraction of $f_{\rm scat}=0.01$, where a small fraction of the intrinsic AGN UV light is observed without dust attenuation. 
            {\it Right}: Excluding all incident UV radiation at $\lambda < 3645~{\rm \AA}$ when computing the dust-reprocessed IR emission (dust-free galaxy component). 
        }
    }
    \vspace{5mm}
    \label{fig:paras_survey_f_scat}
\end{figure*}

\blue{In the former case, we adopt a phenomenological parametrization with a scattering fraction
to account for the possible contribution of scattered/leaked UV radiation from the central AGN \citep{Labbe_2025_LRD_ALMA}.
The parameter $f_{\rm scat}$ represents the fraction of the intrinsic AGN UV flux that reaches the observer without dust attenuation. 
To infer the incident SED from observed flux, we modify Equation~\ref{eqn:de-extinguish} as
\begin{equation}
\label{eqn:de-extinguish_f_scat}
\Lnuinc = \Lnuobs / \left[f_{\rm scat} + (1-f_{\rm scat}) \cdot 10^{-0.4 A_\lambda} \right],
\end{equation}
and repeat the same analysis for A2744-45924 assuming $f_{\rm scat}=0.01$ and SMC opacity law (the left panel of Figure~\ref{fig:paras_survey}). 
As shown in the left panel in Figure~\ref{fig:paras_survey_f_scat}, the overall trend remains similar to our previous results, 
although the upper limit is relaxed to $\AVlim \simeq 1.55~{\rm mag}$. 
This value is still significantly lower than the heavy-dust scenario ($\AVlim \simeq 3.48~{\rm mag}$; Section~\ref{subsec:comparison}).}

\blue{In the latter case, we examine an extreme test case where the rest-frame UV flux is entirely decoupled from the nuclear dusty region, assuming that the observed UV emission originates from the host galaxy without dust attenuation. 
For this purpose, we exclude all spectral data points at wavelengths shorter than the Balmer limit ($3645~{\rm \AA}$) when computing the reprocessed infrared spectrum of heated dust.}

\blue{The right panel of Figure~\ref{fig:paras_survey_f_scat} presents the result of the $A_V$ limit for A2744-45924 under the SMC opacity law. 
With the UV contribution to dust heating removed, the reprocessed infrared spectrum is less constrained and thus the upper limit of extinction increases to $\AVlim \simeq 3.11~{\rm mag}$. 
This least restrictive limit is obtained only for a flat density profile ($\gamma=0$) with a specific density normalization of $n_0 \simeq 55~\cc$. 
However, steeper density profiles ($\gamma \ge 1$), which are physically expected under the BH gravitational potential, yield tighter constraints of $\AVlim \simeq 1.36~{\rm mag}$. }

\blue{It is worth emphasizing that this test excludes any AGN-contributed UV emission underlying below the observed continuum.
This assumption is overly extreme so that the central AGN cannot produce ionizing radiation and thus broad Balmer emission lines.
For instance, if the central AGN is intrinsically dark at $\lambda \leq 3645~{\rm \AA}$, the observed H$\alpha$ luminosity of A2744-45924 
($L_{\rm H\alpha}\simeq 10^{44}~{\rm erg~s}^{-1}$) would originate from the unobscured UV component. 
However, under Case~B recombination, such bright H$\alpha$ emission requires an ionizing UV luminosity of $L_{\rm ion}\simeq 10^{45}~{\rm erg~s}^{-1}$, 
nearly one order of magnitude higher than the observed UV continuum.
This ten-fold discrepancy rules out a purely unobscured stellar origin and instead implies a non-negligible contribution of ionizing UV radiation from the central AGN.}

\bibliography{ms}{}
\bibliographystyle{aasjournalv7}



\end{document}